%% file: main.tex
\documentclass[sigconf]{acmart}

\usepackage{amsmath}
\usepackage[linesnumbered, ruled]{algorithm2e}
\usepackage{booktabs}
\usepackage{multirow}

\renewcommand\footnotetextcopyrightpermission[1]{} 
\begin{document}
\title{Heterogeneous Collaborative Filtering}

\author{Yifang Liu, Zhentao Xu, Cong Hui, Yi Xuan, Jessie Chen, Yuanming Shan}
\thanks{Corresponding author: yifang.liu@smule.com}
\affiliation{%
  \institution{Smule, Inc.}
}

\input{abstract}
\keywords{Recommender, Collaborative Filtering, Algorithm}
\maketitle
\input{introduction}

\input{hcf}
\input{experiment}

\input{conclusion}

\bibliographystyle{abbrv}
\bibliography{paper} 

\end{document}

%% file: abstract.tex
\begin{abstract}
Recommendation system is important to a content sharing/creating social network. Collaborative filtering is a widely-adopted technology in conventional recommenders, which is based on similarity between positively engaged content items involving the same users. Conventional collaborative filtering (CCF) suffers from cold start problem and narrow content diversity. We propose a new recommendation approach, heterogeneous collaborative filtering (HCF) to tackle these challenges at the root, while keeping the strength of collaborative filtering. We present two implementation algorithms of HCF for content recommendation and content dissemination. Experiment results demonstrate that our approach improve the recommendation quality in a real world social network for content creating and sharing.
\end{abstract}

%% file: introduction.tex
\section{Introduction}
\label{sec:introduction}

Content recommendation is to the core of a content creation/sharing social network. A commonly adopted technology in recommenders is collaborative filtering, which has a wide variety of successful algorithm implementations, e.g., FM~\cite{icdm10rendle}, MF~\cite{computer2009koren}, ALS~\cite{jlmr15hastie}, SVD++~\cite{kdd08koren}, PITF~\cite{wsdm10rendle}, FPMC~\cite{www10rendle}. Conventional collaborative filtering (CCF), such as the aforementioned algorithms, often adopts homogeneous inferences. In this paper, we call them homogeneous collaborative filtering, because the recommenders predict future positive engagements (i.e., the user is interested in the presenting items) based on the user's past positive engagements. The inference is homogeneous, in the sense that the recommended items and the past engagements are "similar" to each other in the same direction of engagement. As the most common recommendation approach, positive-to-positive inference reflects the co-occurrence of items engaged by the same users. 

By the same reasoning, negative-to-negative inference - the other form of homogeneous collaborative filtering - can be used to predict/infer future negative engagement for a user (i.e., the user shows no interest in the presenting items). For example, based on the items that received a "Thumb-Down" rating from the user in the past, "similar" items are excluded from future recommendation for that user. In this paper, conventional collaborative filtering (CCF) and homogeneous collaborative filtering are two interchangeable terms.

One of the main strengths of the conventional collaborative filtering (CCF) is high effectiveness and precise prediction of positive engagements. The effectiveness comes with an important premise - sufficient engagement data must be available for model training. This condition is the root cause of a major challenge faced by CCF - \textbf{cold start}. When a user is new to a content creating/sharing social network, the recommender knows little about the user, especially about the user's positive engagement on any item. The sparse data makes it hard for CCF to come up with reasonably relevant content recommendations. The same reasoning can be applied to content cold start, where CCF is ineffective in figuring out which users would potentially be interested in a new content item with little engagement data.

CCF faces another challenge in positive engagement prediction - \textbf{content diversity}: over time, the recommendation becomes more and more focused on what the user has shown strong interest in. While the positive engagement prediction becoming more and more precise with more positive engagement data accumulates, the scope of predicted user interest becomes narrower and narrower. The converging recommendation scope affects the content diversity for each user, in general. Consequently, the engagement feedback loop eliminates/excludes a large space of potentially very interesting contents for the user. In other words, on the balance spectrum of exploitation vs. exploration, CCF tends to weigh mostly on exploitation. 

CCF can try different fixes to overcome the two major challenges above, but resolutions' effectiveness will be fundamentally limited by the homogeneous-inference nature of CCF. To tackle this challenge at its core, we propose \textbf{Heterogeneous Collaborative Filtering (HCF)}. 

HCF employs heterogeneous inference, which takes two forms: negative-to-positive inference, positive-to-negative inference. In this paper, we focus on the former; any future mentions of HCF is solely on the aspect of negative-to-positive inference. More specifically, negative items that were not interesting to a user in the past are used to infer positive items that are likely to be of interest to the user. 
Essentially, HCF leverages negative correlation between items, in inferring future positive item engagements based on past negative item engagements. 
In a loose statistical sense, "similarity" in collaborative filtering acts like a proxy for "correlation" between entities. CCF only cares about positive correlation between entities. HCF complements CCF with negative correlation.

Note that HCF can be used in two recommendation directions: content recommendation - suggesting items to a user, content dissemination - suggesting users to an item: inferring positive users who may be interested in the item, based on negative users who are not interested in an item in the past. We call the later a content dissemination model, while the former is considered as the standard content recommendation engine. 
HCF can be equally-readily applied in both directions, because both are essentially suggesting relevant candidate entities to a target entity in interest. The difference is: the candidate entities in content recommendation are content items (relevant to a target user), while the candidate entities in content dissemination are impressible users (targeting a specific item).

HCF's ability to mitigate the cold-start problem originates in its philosophy of leveraging the usually more abundant negative engagement data. 
HCF's potential to address the shrinking content diversity problem is attributed to the divergent nature of negative-to-positive inference. Basically, there are many different possible positives given one negative. 

\textbf{The main contributions} of this paper include:
\vspace{-0.06in}
\begin{enumerate}
\item We present a new approach of recommendation - Heterogeneous Collaborative Filtering (HCF).
\item We present HCF's application for content recommendation to address the challenges of cold start and content diversity.
\item We present HCF's application on content dissemination, which is a novel approach for recommendation, on its own.
\item We implement the HCF approach in a real world social network - Smule - and test it with real production data.
\end{enumerate}

%% file: hcf.tex
\section{HCF}
\label{sec:hcf}

The core idea of heterogeneous collaborative filtering (HCF) is to predict/infer positive engagements, based on both homogeneous and heterogeneous similarity between entities. It is important to note that HCF does not exclude CCF. In fact, HCF usually employs a combination of heterogeneous and homogeneous inference, and it even encourages modeling the interaction between the two.

The rest of this section will first explain a high-level framework of HCF approach, followed by the presentation of a few exemplary algorithm implementations of HCF for a real-world social network application. The exemplary algorithms illustrate how HCF can be applied to two forms of application: content recommendation and content dissemination. 

Before diving into the general HCF framework, we need to clarify the terms of negative engagement and positive engagement. An engagement means the response and depth of the user's involvement with a content item, which can be either implicit or explicit. More specifically, examples of implicit positive and negative engagements include selection/click and non-selection/ignoring of a content item displayed to a user, respectively; examples of explicit positive and negative engagements include users' thumb-up and thumb-down on an item, respectively. 

In general, HCF usually works in the following high-level steps on a recommendation/dissemination problem:

Step 1) Obtain the candidate set: the candidate items to recommend to a user, or the candidate users to expose an item to. The candidate set can be the whole population available, or a subset of the population that is selected by heterogeneous and homogeneous similarity between candidates.

Step 2) Use a scoring algorithm, combining heterogeneous and homogeneous inferences (preferably with the modeling of the interaction between the two types of inference), to score and rank individual candidates, and then choose the top candidates as the recommendation. The output score of the scoring algorithm indicates the likelihood of the user selecting/engaging with the item. The input variables of the algorithm include: an entity pair consisting of the target entity and the candidate entity, their properties, together with a bank of positive and negative engagement instances over all relevant users and items, including the user and the item in interest. Which of the two entities is a user or an item depends on the model being for content recommendation or for content dissemination.

The main benefit of pre-selecting candidates in step 1 of the HCF framework ensures a diverse coverage of inputs to the scoring algorithm in step 2, especially when a large percentage of candidates is guaranteed to enter the final recommendation list.

The candidate selection can be done by choosing the entities that are most "similar" (heterogeneously or homogeneously) to the entities in the past engagements. The Cosine similarity between the characteristic vectors of two entities is defined in Equ.~\ref{equ:cosine-sim}. 

\vspace{-0.15in}
\begin{align} \label{equ:cosine-sim}
& \mbox{similarity} (\mathbf{u}_i, \mathbf{u}_j) = \frac{\langle \mathbf{u}_i, \mathbf{u}_j \rangle }{||\mathbf{u}_i|| \mbox{\ } ||\mathbf{u}_j||}
\end{align}
where $\langle \mathbf{u}_i, \mathbf{u}_j \rangle$ denotes the dot product of the characteristic vectors $\mathbf{u}_i$ and $\mathbf{u}_j$ of entity $i$ and $j$; the characteristic vector $\mathbf{u}_i$ consists of engagement instances involving entity $i$, $[e_{i,1}, e_{i,2}, ..., e_{i,n}]$. In the case of content recommendation, the entities in Equ.~\ref{equ:cosine-sim} represent content items, and the engagement instances in its characteristic vector $\mathbf{u}_i$ indicates users who have engaged with it; in the case of content dissemination, the entities in Equ.~\ref{equ:cosine-sim} represent users, and the engagement instances in $\mathbf{u}_i$ indicate items that the user has engaged with.  

The engagement instances in a characteristic vector must be either all positive or all negative. When $\mathbf{u}_i$ and $\mathbf{u}_j$ being both positive, Equ.~\ref{equ:cosine-sim} gives the homogeneous similarity between entities $i$ and $j$; otherwise if $\mathbf{u}_i$ and $\mathbf{u}_j$ being one positive and one negative, it gives the heterogeneous similarity between the two. 
The candidate selection combines two subsets of candidates calculated by Equ.~\ref{equ:cosine-sim} for heterogeneous similarity and homogeneous similarity, respectively. The union of the two subsets forms the overall candidate set. 

In step 2 of the HCF framework, modeling the interaction between HCF and CCF can be done in various ways. For example, GBDT (gradient boosted decision trees) or FM (factorization machines) can be used as the model to drive the scoring algorithm. For the sake of demonstration, this paper chooses to explain a 2-way $k$-factor FM model in Equ.~\ref{equ:fm-hcf} as the scoring model. For the simplicity of presentation, here the input variables to the FM model only include the user and the item in interest, as well as a bank of positive and negative engagement instances among all relevant users and items. That is, we ignore other potentially very informative inputs, such as the properties of the given user and item.  

\vspace{-0.15in}
\begin{align} \label{equ:fm-hcf}
& \hat{y} (\mathbf{x}) = w_0 + \sum_{i=1}^{n} w_i x_i + \sum_{i=1}^{n} \sum_{j=i+1}^{n} \langle \mathbf{v}_i, \mathbf{v}_j \rangle x_i x_j 
\end{align}
where $x_i$ denotes the indicator variable for the $i$-th active entity (the entity can be either an item or a user that is involved in a positive or negative engagement/impression); $\mathbf{v}_i \in \mathbb{R}^{k}$ represents the k-factor vector of the $i$-th entity; the dot product $\langle \mathbf{v}_i, \mathbf{v}_j \rangle$ indicates the interaction between the $i$-th and the $j$-th entity; $\hat{y}$ is the output score - the predicted likelihood of the positive engagement involving the user and the item. The model parameters need to be fitted include: $w_0, \mathbf{w} \in \mathbb{R}^{n}, \mathbf{v} \in \mathbb{R}^{n \times k}$.

When $x_i$ and $x_j$ indicate one positive engagement and one negative engagement, their heterogeneous-homogeneous interaction is captured by the term $\langle \mathbf{v}_i, \mathbf{v}_j \rangle$; when $x_i$ and $x_j$ indicate both positive engagements or both negative engagements, the dot product models the interaction between homogeneous engagements as in conventional collaborative filtering.

The FM model in Equ.~\ref{equ:fm-hcf} can be applied to both content recommendation and content dissemination. When each of multiple candidate items is paired with a given user for scoring by Equ.~\ref{equ:fm-hcf}, the model predicts the engagement likelihood for content recommendation; otherwise if each of multiple candidate users is paired with a given item for scoring by Equ.~\ref{equ:fm-hcf}, the model predicts the engagement likelihood for content dissemination. 


Next, we look at how HCF framework can be applied to content recommendation and content dissemination, respectively.

\subsection{HCF content recommendation}

This section discusses how to apply HCF to the content recommendation problem - deciding which items to recommend to a specific user. In the exemplary HCF algorithm, the candidate set is constructed as a combination of heterogeneously-similar items and homogeneously-similar items, and use FM model as the scoring function. Alg.~\ref{alg:hcf-item2item} outlines the exemplary algorithm of HCF for content recommendation. 

\begin{algorithm}[t]
\SetKwData{Left}{left}
\SetKwData{This}{this}
\SetKwData{Up}{up}
\SetKwFunction{Union}{Union}
\SetKwFunction{FindCompress}{FindCompress}
\SetKwInOut{Input}{input}
\SetKwInOut{Output}{output}
\SetKwInOut{Parameter}{parameter}
\SetKwRepeat{Do}{do}{while}
\Input{the target user to receive suggested contents; all available content items; historical engagement instances over all items and all users.}
\Output{the engagement likelihood scores for items to be recommended to the target user.}
\BlankLine
    Step 1) construct the candidate item set by combining two subsets calculated by Equ.~\ref{equ:cosine-sim}: one from heterogeneously-similar items, one from homogeneously-similar items\;
    Step 2) score the candidate items by the FM model in Equ.~\ref{equ:fm-hcf}, where the candidate entities are content items\;
\caption{HCF CONTENT RECOMMENDATION} \label{alg:hcf-item2item}
\end{algorithm}

\subsection{HCF content dissemination}

This section discusses an application of HCF to content dissemination - deciding which users to expose to a specific item. In an iterative content dissemination process, a group of users are chosen to be presented with a given item in each iteration, based on previous users' response to the item in previous iterations. A heuristic for random user selection is employed by Tik Tok's so-called decentralized video recommendation. 

From a recommendation perspective, HCF content dissemination can be viewed as a process of finding users who could be interested in a given content item. Content dissemination has two goals: evaluating the quality of a content item, and in the case of confirmed item quality, spreading the high-quality items to potentially interested users. For a newly introduced content item with little engagement data, both the quality validation and audience targeting are big challenges for conventional content recommendation approaches. 

HCF for content dissemination takes an unconventional route - leveraging both positive and negative engagements to effectively figure out the next group of users who are potentially interested in the item, and thus help to validate the item's quality. Instead of finding relevant items for a user, the hope is that the right audience can be picked for the item by the heterogeneous inference logic. In other words, HCF content dissemination introduces two novel ideas: 1) The potential audience to a target item is chosen based on the similarity between users. This is expected to be more efficient than the conventional approach - use content recommendation for content spreading, i.e., a content recommendation engine lets the target item take chances to be picked as a recommended item for some users, who may happen to be the right audience in the next iteration of spreading the target item. 2) Heterogeneous inference (complementing CCF) plays a critical role in efficiently identifying the most relevant audience for a new content item. Besides choosing users homogeneously-similar to those who were previously positive on the target item, HCF also picks users heterogeneously-similar to those who were previously negative on the target item. It benefits from the rich information in negative engagements, as well as the interaction between negative and positive engagements. 

In the following exemplary HCF algorithm, we consider how to select a group of users in each iteration of an iterative content spreading process. The candidate set preparation is done by combining heterogeneously-similar users and homogeneously-similar users. The scoring step is done based on the FM model, where the candidate entities are impressible users. Alg.~\ref{alg:hcf-user2user} outlines the exemplary algorithm of HCF for content dissemination. 

\begin{algorithm}[t]
\SetKwData{Left}{left}
\SetKwData{This}{this}
\SetKwData{Up}{up}
\SetKwFunction{Union}{Union}
\SetKwFunction{FindCompress}{FindCompress}
\SetKwInOut{Input}{input}
\SetKwInOut{Output}{output}
\SetKwInOut{Parameter}{parameter}
\SetKwRepeat{Do}{do}{while}
\Input{the target item to be spread among users; all available users; historical engagement instances over all items and all users.}
\Output{the engagement likelihood scores of suggested users to be exposed to the target item.}
\BlankLine
    Step 1) construct the candidate user set by combining two subsets calculated by Equ.~\ref{equ:cosine-sim}: one from heterogeneously-similar users, one from homogeneously-similar users\;
    Step 2) score the candidate users by the FM model in Equ.~\ref{equ:fm-hcf}, where the candidate entities are users\;
\caption{HCF CONTENT DISSEMINATION} \label{alg:hcf-user2user}
\end{algorithm}

%% file: experiment.tex
\section{Experiment}
\label{sec:experiment}

This section explains our evaluation of HCF for its applications on content recommendation and content dissemination, in a real-world social network - Smule. We compare HCF models (the proposed approach) with CCF models (the baseline) on the ROC AUC. 

For the interest of privacy protection, all data in the representation of the experiments, including the input data and the readout results, are provided in relative terms. The true input data used in the experiments are content engagement events involving real users from all over the world, i.e., on the whole user population over all available content items in the real-world production of Smule.  

Model validations are performed offline on historical data. Additional A/B tests are done in Smule's production environment for bench-marking against baseline metrics.

\subsection{Problems in the experiments}

The application of HCF is demonstrated by three exemplary problems listed below.

\textbf{Content recommendation for all users (RecoAll)}: this is standard recommendation problem.

\textbf{Content recommendation for new users (RecoNew)}: this is standard recommendation problem restrained on new users - users who joined the Smule social network within a short period of time.
 
\textbf{Content dissemination for new items (DismNew)}: this is the problem of exposing a new content item (which was uploaded to the network within a short period of time) to the most relevant users for quality validation and popularity promotion. The spreading of a new content item is an iterative process. In each iteration, a new group of users are selected to be presented with the item, based on the response (positive or negative engagement with the item) from the previous users in the previous iterations. The new group of users may grow or shrink in size, depending on the result of the quality validation.  

It is important to note that we only had time to A/B test the first two problems in an online production environment, while the last problem only went through the offline AUC evaluation. We will explain how missing online testing affects the model training data balance, and hence affects how much the model improvement can be reflected in offline AUC evaluation. 
 
\subsection{Algorithms in the experiments}
\label{sec:algo-in-eval}

Following are proposed algorithms and baseline algorithms for each problem in our experiment. 

\textbf{RecoAll}: baseline: CCF with GBDT; proposed: HCF with GBDT

\textbf{RecoNew}: baseline: CCF with FM; proposed: HCF with FM

\textbf{DismNew}: baseline: CCF with FM; proposed: HCF with FM

All the algorithms in the experiment run on the same computer, which is equipped with a 8-core processor and 128GB memory. No GPU computation is involved in the experiment.

\subsection{Results and observations}

In our experiments, the model for every algorithm in all three problems is trained every day, on Smule production data at corresponding times. 

\begin{figure}[t!] \centering
\includegraphics[width=2.1in,angle=0]{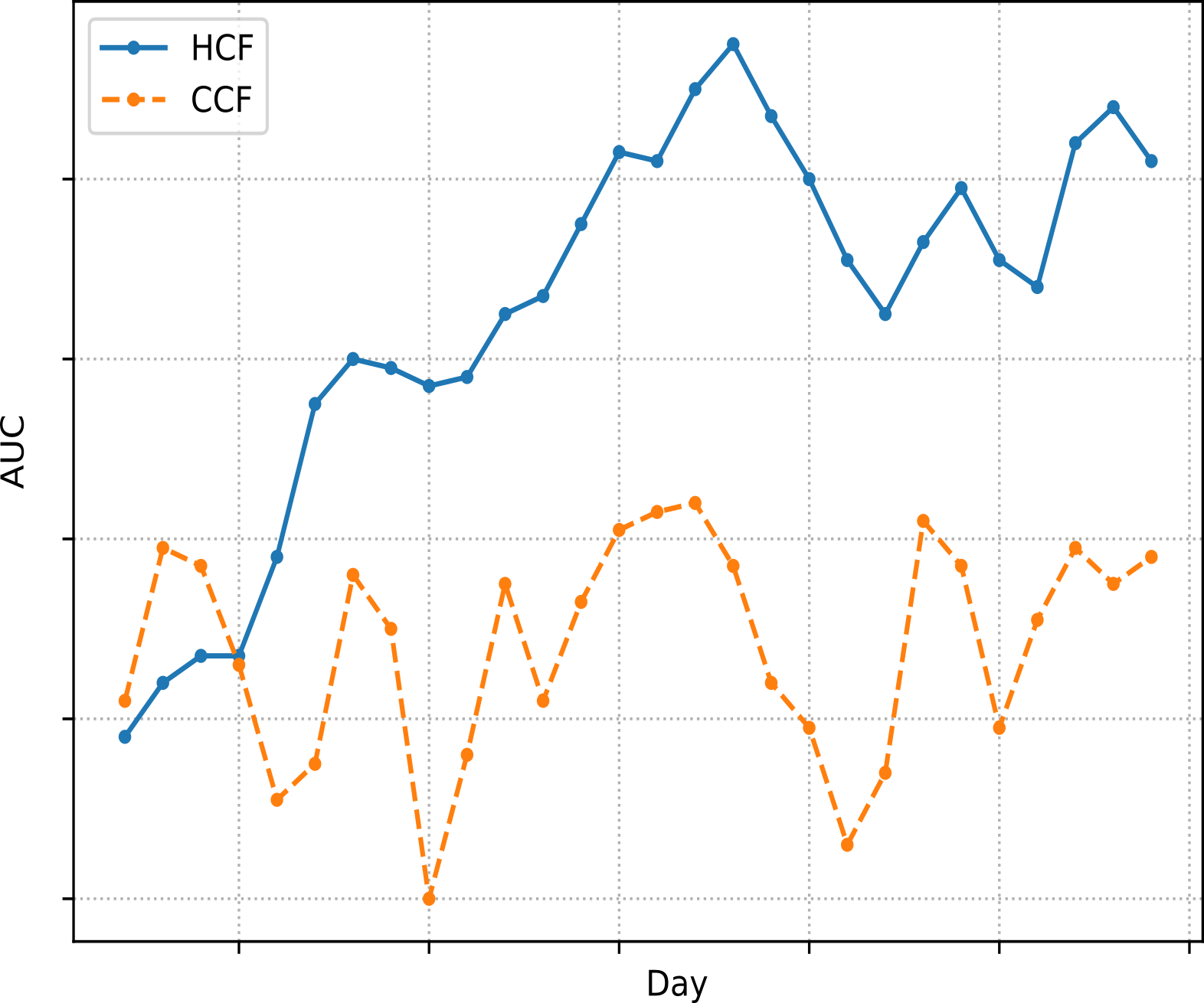}
\caption{AUC before and after the A/B test starts}
\vspace{-0.1in}
\label{fig:auc-change}
\end{figure}

Fig.~\ref{fig:auc-change} shows the AUC changes of the CCF algorithm and the HCF algorithm on the first problem - content recommendation for all users. The time span of the AUC measurement starts from two days before the test begins, and concludes 20+ days into the test. By the way, the positive result of the A/B test became statistically significant on the first day of the test. 

It is clear that the AUC improvement by our HCF algorithm only begins to show after the A/B test starts. This is because without the HCF algorithm injecting negatively correlated data into the production, the training data (which is historical production data) does not contain the necessary information to support the heterogeneous inference. This effect of missing information in the training data will show its impact on model quality evaluation again in the last experiment problem, since the last problem has not yet been tested online in production.   

The relative AUC improvement on all three problems are listed in Table~\ref{tab:auc-results}. Our offline and online experiment results show that in these problems, $10\%$ improvement on offline AUC roughly translates into $25-35\%$ improvement on online engagement KEIs, e.g., view-through rate.

The fact that HCF is unlikely to improve the recommendation relevance over CCF implies that the AUC improvement by HCF on the first problem attributes mostly to HCF's advantage on content diversity. On the other hand, a large portion of the AUC improvement by HCF on the second problem can be the result of HCF's cold start resolving ability, since recommendation relevance matters less to new users. On the third problem, the full potential of HCF does not show up in our offline AUC evaluation (without online test). This is expected, for the same reason as the AUC change in the first problem. 

The runtimes of all algorithms are listed in Table~\ref{tab:alg-runtime}.

\begin{table}[ht] \centering
	\scriptsize
	\caption{AUC improvement (HCF AUC relative to CCF AUC)}
	\label{tab:auc-results}
	\begin{tabular}{| l | r | r |}
		\hline
		The problem & Baseline algo (CCF)   & Proposed algo (HCF) \\
		\hline
		RecoAll     & $100\%$                 & $110.2\%  \uparrow 10.2\%$ \\ 
		\hline
		RecoNew     & $100\%$                & $118.6\%  \uparrow 18.6\%$ \\ 
		\hline
		DismNew     & $100\%$              & $102.5\%  \uparrow 2.5\%$  \\ 
		\hline
	\end{tabular}
\end{table}

\begin{table}[ht] \centering
	\scriptsize
	\caption{Algorithm runtime (seconds)}
	\label{tab:alg-runtime}
	\begin{tabular}{| l | r | r |}
		\hline
		The problem & Baseline algo (CCF)   & Proposed algo (HCF) \\
		\hline
		RecoAll     & 2476                 & $2599 \uparrow 5\%$ \\ 
		\hline
		RecoNew     & 3.29                 & $12.55 \uparrow 281\%$  \\ 
		\hline
		DismNew     & 181                  & $355 \uparrow96\%$ \\ 
		\hline
	\end{tabular}
\end{table}

%% file: conclusion.tex
\section{Conclusion}
\label{sec:conclusion}

In this paper, we proposed two novel approaches for recommendation: 1) heterogeneous collaborative filtering (HCF). 2) HCF for content dissemination. They leverage heterogeneous similarity between content items or impressible users in their relevance inference. The proposed techniques yield notable improvements on recommendation quality in our experiments on production data from a real-world social network - Smule. The evaluation suggests that the techniques are effective in tackling content diversity challenge and cold start challenge.